\documentclass[sigconf, nonacm]{acmart}

\usepackage{array}
\usepackage{multirow}
\usepackage{xspace}
\usepackage{makecell}
\usepackage{algorithm}
\usepackage{algpseudocode}

\usepackage{enumitem}
\setlist[itemize]{
  leftmargin=1.2em,
  labelsep=0.4em
}
\setlist[enumerate]{
  leftmargin=1.5em,
  labelsep=0.4em
}

\usepackage{etoolbox}

\AtBeginEnvironment{equation}{\small}
\AtBeginEnvironment{equation*}{\small}
\AtBeginEnvironment{align}{\small}
\AtBeginEnvironment{align*}{\small}
\AtBeginEnvironment{gather}{\small}
\AtBeginEnvironment{gather*}{\small}
\AtBeginEnvironment{multline}{\small}
\AtBeginEnvironment{multline*}{\small}

\newcommand{\oursname}{Bridge}

\DeclareRobustCommand{\ours}{\textsc{\oursname}\xspace}

\newcommand{\sphinteract}{\textsc{Sphinteract}\xspace}

\newcommand{\seedname}{Seed}
\DeclareRobustCommand{\seed}{\textsc{\seedname}\xspace}

\newcommand{\seedmodel}{\textsc{Qwen3-Coder-30B-A3B-Instruct}\xspace}
\newcommand{\onlinemodel}{\textsc{Qwen3.5-27B}\xspace}
\newcommand{\offlinemodel}{\texttt{qwen3.7-plus}\xspace}
\newcommand{\embeddingmodel}{\textsc{Qwen3-Embedding-0.6B}\xspace}

\algrenewcommand\algorithmicrequire{\textbf{Input:}}
\algrenewcommand\algorithmicensure{\textbf{Output:}}

\newcommand\vldbdoi{XX.XX/XXX.XX}
\newcommand\vldbpages{XXX-XXX}

\newcommand\vldbvolume{14}
\newcommand\vldbissue{1}
\newcommand\vldbyear{2020}

\newcommand\vldbauthors{\authors}
\newcommand\vldbtitle{\shorttitle}
\newcommand\vldbavailabilityurl{https://github.com/slyipae1/BRIDGE.git}
\newcommand\vldbpagestyle{plain}

\begin{document}

\title{\ours: Database-Grounded Interactive Disambiguation for Text-to-SQL}

\author{Sau Lai Yip}
\affiliation{%
  \institution{The Hong Kong University of Science and Technology}
  \city{Hong Kong}
  \country{China}
}
\email{slyipae@connect.ust.hk}

\author{Yue Pang}
\affiliation{%
  \institution{The Hong Kong University of Science and Technology}
  \city{Hong Kong}
  \country{China}
}
\email{michellepy@ust.hk}

\author{Shiwen Wu}
\affiliation{%
  \institution{The Hong Kong University of Science and Technology}
  \city{Hong Kong}
  \country{China}
}
\email{swubs@connect.ust.hk}

\author{Xiaofang Zhou}
\affiliation{%
  \institution{The Hong Kong University of Science and Technology}
  \city{Hong Kong}
  \country{China}
}
\email{zxf@cse.ust.hk}

\begin{abstract}
Ambiguity remains a persistent challenge in text-to-SQL, where an underspecified natural-language question may admit multiple plausible SQL interpretations.
Recent work attempts to recover the user's intended interpretation through iterative clarification, but typically relies heavily on LLM parametric knowledge for ad hoc reasoning, leading to limited ambiguity-resolution effectiveness and high LLM cost.
We further observe that many ambiguity errors stem from ambiguous column grounding, which remains poorly handled.
Therefore, we propose \ours, a database-grounded disambiguation framework built on an existing iterative protocol. \ours supplies the LLM's ambiguity detection with offline-distilled knowledge about potential ambiguity relations among columns in the target database.
Offline, \ours constructs a \emph{Column Ambiguity Graph} by first identifying potentially ambiguous column pairs using complementary database-grounded signals, followed by LLM-based verification.
Online, columns and values mentioned in a previous incorrect SQL prediction are used as anchors to retrieve alternatives from the Column Ambiguity Graph and an offline-constructed value similarity index, respectively. The results are integrated for further pruning and consolidated for subsequent ambiguity detection.
Extensive experiments on BIRD-dev without the benchmark-provided evidence show that, at comparable LLM cost, \ours outperforms the strongest baseline by 11.60\% in execution accuracy and improves the recall and precision of column usage in the final SQL by 5.25\% and 12.37\%, respectively.
\end{abstract}

\maketitle

\pagestyle{\vldbpagestyle}
\begingroup\small\noindent\raggedright\textbf{PVLDB Reference Format:}\\
\vldbauthors. \vldbtitle. PVLDB, \vldbvolume(\vldbissue): \vldbpages, \vldbyear.\\
\href{https://doi.org/\vldbdoi}{doi:\vldbdoi}
\endgroup
\begingroup
\renewcommand\thefootnote{}\footnote{\noindent
This work is licensed under the Creative Commons BY-NC-ND 4.0 International License. Visit \url{https://creativecommons.org/licenses/by-nc-nd/4.0/} to view a copy of this license. For any use beyond those covered by this license, obtain permission by emailing \href{mailto:info@vldb.org}{info@vldb.org}. Copyright is held by the owner/author(s). Publication rights licensed to the VLDB Endowment. \\
\raggedright Proceedings of the VLDB Endowment, Vol. \vldbvolume, No. \vldbissue\ %
ISSN 2150-8097. \\
\href{https://doi.org/\vldbdoi}{doi:\vldbdoi} \\
}\addtocounter{footnote}{-1}\endgroup

\ifdefempty{\vldbavailabilityurl}{}{
\vspace{.3cm}
\begingroup\small\noindent\raggedright\textbf{PVLDB Artifact Availability:}\\
The source code, data, and/or other artifacts have been made available at \url{\vldbavailabilityurl}.
\endgroup
}

\begin{figure*}[t]
    \centering
    \includegraphics[width=\linewidth]{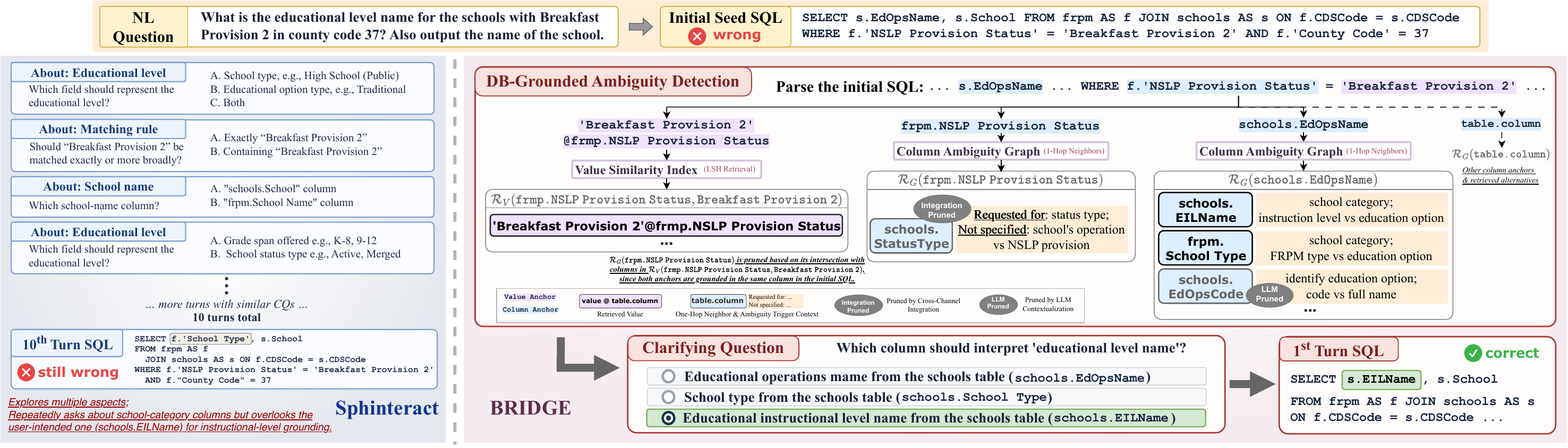}
    \caption{Running example from BIRD-dev~\cite{BIRD}. \sphinteract~\cite{sphinteract} fails to identify the intended column for ``educational level name'' after 10 turns, while \ours retrieves relevant ambiguity relations and resolves the key issue in the first turn.}
    \label{fig:running-example}
\end{figure*}

\section{Introduction}
\label{sec:introduction}

Text-to-SQL enables users to query relational databases using natural language. While large language models (LLMs) have substantially advanced performance on benchmarks such as Spider~\cite{Spider} and BIRD~\cite{BIRD}, stronger generation capabilities alone cannot resolve a more fundamental dilemma: real-world user requests can be inherently underspecified, admitting multiple plausible SQL interpretations under the same database. Such issues are prevalent in practice: Jin et al.\ attribute 29.66\% of audited BIRD Mini-Dev annotation errors and 23.68\% of Spider 2.0-Snow errors to ambiguity~\cite{benchmarkBroken}; Zhao et al.\ further show that SQL queries written independently by professionals often produce different answers~\cite{sphinteract}. Therefore, a reliable system must be able to identify ambiguous interpretations and request clarification to determine the user's intent.

Existing work has explored various ways to address ambiguity in Text-to-SQL. Earlier approaches primarily focus on detecting ambiguity and exposing multiple plausible SQL interpretations~\cite{KnowWhatIDontKnow,AmbiQT,DisambiguateFirst,ODIN}, placing a substantial burden on users to identify the intended one among the alternatives. More recent work increasingly seeks to resolve ambiguity to a single interpretation, with interactive methods that obtain missing information by asking users becoming the predominant direction. Their ambiguity-detection mechanisms mainly follow two strategies: (1) \emph{uncertainty-based} methods that quantify model uncertainty during SQL prediction~\cite{ReliableText2SQL,AtomSQL}; and (2) \emph{taxonomy-based} methods that define ambiguity types to guide the detection~\cite{clear,AmbiSQL,sphinteract}. The detected issues are then formalized as clarification questions (CQs). Despite these mechanisms, existing approaches typically rely heavily on ad hoc reasoning based solely on LLM parametric knowledge, without sufficient grounding in the target database, limiting both the effectiveness and efficiency of ambiguity resolution.

Furthermore, many ambiguity errors are associated with ambiguous column grounding, which remain challenging for existing methods~\cite{CLARITY}. We observe that on BIRD-dev~\cite{BIRD}, \sphinteract~\cite{sphinteract}, a state-of-the-art method that asks CQs and regenerates SQL in multiple turns, has 70.05\% of its initial failure cases stemming from incorrect column usage. Even after 10 interaction turns, 56.56\% of these cases remain incorrect with wrong column usage, accounting for 62.67\% of all remaining failures. This suggests the significance of addressing column-grounding ambiguity, although these figures should be viewed as an upper bound because not every wrong-column failure necessarily results from ambiguity. 

To address these limitations, we propose to construct a \emph{Column Ambiguity Graph} to achieve database-grounded disambiguation. Specifically, we distill potential ambiguity relations among database columns into a compact graph offline, and retrieve only the query-relevant subset to serve as online reasoning cues. This combination of offline distillation and query-specific retrieval strengthens ambiguity detection while reducing query-time reasoning cost.

The example in Figure~\ref{fig:running-example} illustrates this limitation. The request asks for the ``educational level name'' of schools satisfying several conditions, while the initial SQL does not meet the user's intent for grounding this phrase to \texttt{schools.EdOpsName}. \sphinteract~\cite{sphinteract} explores different conditions of the request. Although it revisits the educational-level ambiguity repeatedly, considering different related school-category columns, it never exposes the intended \texttt{schools.EILName} and the ambiguity remained unresolved after 10 turns. 
In contrast, \ours retrieves relevant offline-distilled ambiguity relations, successfully guiding the LLM to ask about the intended column and resolve the issue with only one iteration turn.
This behavior illustrates the difficulty of relying primarily on LLM parametric knowledge to discover concrete competing database groundings at query time, as the model may blindly explore plausible directions while missing some critical alternative.

Building and effectively using the Column Ambiguity Graph presents two main challenges.

\textbf{Challenge 1: How can useful column ambiguity relations be effectively distilled into the graph?}
Recording all or arbitrary column relations would introduce substantial noise, since relatedness does not necessarily imply a plausible ambiguity. The graph should therefore retain only column pairs that can realistically become competing interpretations. Moreover, storing the column pair alone provides limited guidance for reliable query-time reasoning; the graph must also distill sufficient information about why and under what conditions the pair may become ambiguous.

\textbf{Challenge 2: How can only the query-relevant subset be retrieved with high precision?}
A user question typically involves only a small portion of the database. Supplying the entire graph would increase both token cost and irrelevant context, while overly restrictive retrieval risks missing critical alternatives. Designing an effective retrieval mechanism that balances these two concerns is therefore essential for reliable and efficient ambiguity detection.

We address these challenges with \ours, a database-grounded ambiguity-resolution framework that inserts a front-loaded clarification stage into the state-of-the-art iterative post-generation interaction protocol~\cite{sphinteract}. \ours separates database-level ambiguity induction from query-specific detection:

\textbf{Offline.} \ours constructs the \emph{Column Ambiguity Graph} with two steps: \emph{Multi-View Candidate Proposal} identifies potentially ambiguous column pairs from three complementary signals, then \emph{LLM Induction and Verification} verifies these pairs and summarizes ambiguity reasons. Additionally, a value similarity index is built to support value-based retrieval at query time for a complementary source of information.
    
\textbf{Online.} Given an initially incorrect SQL prediction, \emph{Dual-Channel Retrieval} uses the columns and values mentioned in the SQL as anchors to retrieve query-relevant column alternatives from the graph and value alternatives from the value index. \emph{Cross-Channel Integration} then associates and filters the results from the two sources, which then serve as reasoning cues for the LLM to identify unresolved ambiguity and formulate CQs.

We summarize our contributions as follows:
\begin{itemize}

    \item We identify column-grounding ambiguity as a significant unresolved issue and propose \ours, a database-grounded disambiguation framework built around a \emph{Column Ambiguity Graph}.
    
    \item We design an offline construction pipeline that identifies potentially ambiguous column pairs from complementary signals, verifies them with an LLM, and summarizes their ambiguity reason into reusable graph annotations.
    
    \item We design an online ambiguity-detection procedure that retrieves query-relevant column and value alternatives, integrates the two sources, and uses the resulting information to guide clarification-question generation.
    
    \item Extensive experiments is performed on BIRD-dev~\cite{BIRD} without BIRD-provided evidence for a more challenging setting. At comparable LLM cost, \ours outperforms the strongest baseline~\cite{sphinteract} in accuracy by 11.60\%, reaching 77.38\%, and in column-linking recall and precision by 5.25\% and 12.37\%, reaching 85.68\% and 93.96\%, respectively.
\end{itemize}

\section{Related Work}
\label{sec:related-work}

\paragraph{Text-to-SQL}
Text-to-SQL has evolved substantially with the emergence of LLMs, which are now widely used in recent systems. LLM-based approaches broadly follow two lines. \emph{Fine-tuning-based} approaches specialize LLMs for Text-to-SQL through SQL-centric supervised fine-tuning~\cite{CodeS,OmniSQL,XiYanSQL}, with recent work further employing reinforcement learning to improve complex SQL reasoning~\cite{SQLR1,RewardSQL}. \emph{Prompting-based} approaches instead leverage general-purpose LLMs without task-specific training, often decomposing Text-to-SQL into multiple stages such as schema linking, SQL generation, refinement, and candidate selection~\cite{DAILSQL,DINSQL,CHESS,OpenSearchSQL,DeepEyeSQL,DPC}. These advances strengthen the underlying Text-to-SQL capability through either fine-tuning or inference-time orchestration. 

Our work is complementary: rather than modifying how the base parser is trained or organized, we focus on ambiguity in underspecified requests, identifying competing database-grounded interpretations and eliciting the intended one for SQL generation.



\paragraph{Ambiguity Detection \& Resolution in Text-to-SQL}
Ambiguity is a critical challenge in Text-to-SQL, where one NL question may admit multiple plausible SQL interpretations. 
Earlier work primarily focuses on detecting ambiguity without determining which one a particular user intends~\cite{KnowWhatIDontKnow,AmbiQT,DisambiguateFirst, ODIN}. These approaches typically return multiple SQLs, but ultimately require the user to distinguish among the alternatives to recover the intended one, imposing additional cognitive burden, especially on non-proficient users.
More recent work aims at obtaining a single resolved interpretation. One line pursues \emph{automatic} resolution: \textsc{SOMA-SQL}~\cite{SOMASQL} infers the most likely interpretation from patterns in historical query logs and execution probing. 
\emph{Interactive} approaches, the predominant direction, instead obtain the missing information directly from users through clarification. These methods mainly follow two strategies. 
\emph{Uncertainty-based} approaches identify ambiguity from model uncertainty: \textsc {RTS}~\cite{ReliableText2SQL} quantifies uncertainty during schema linking to trigger abstention and human confirmation, while \textsc{AtomSQL}~\cite{AtomSQL} uses clause-level model disagreement to identify uncertain and elicit user preferences among competing alternatives for SQL clauses. 
More works instead use \emph{taxonomy-based} approaches, which define ambiguity types to guide the formulation of CQs, as in \textsc{CLEAR}~\cite{clear}, \textsc{AmbiSQL}~\cite{AmbiSQL}, and \sphinteract~\cite{sphinteract}.

Despite these mechanisms, current approaches solely rely on LLM parametric knowledge and ad hoc reasoning to identify competing interpretations, limiting both ambiguity-resolution effectiveness and efficiency. Our work instead supplies ambiguity detection with competing column relations.

\section{Preliminaries}
\label{sec:preliminaries}

\subsection{Problem Formulation}
\label{sec:problem-formulation}

In this section, we formally define the Text-to-SQL task and the ambiguity problem. We then provide a formulation of interactive disambiguation for this problem, which is the focus of this work.

\paragraph{Text-to-SQL}
The Text-to-SQL task aims to translate a natural-language (NL) question into a SQL query that correctly reflects the user's intent.
Formally, given an NL question $nl$ with the user's underlying intended interpretation $I^\star$ that is not explicitly specified and a database instance $D$, the objective is to produce a prediction $\hat q$ satisfying
$\operatorname{Exec}(\hat q,D) = \operatorname{Exec}_{\mathrm{intent}}(I^\star,D)$,
where $\operatorname{Exec}(q,D)$ denotes the execution result of $q$ on $D$, and $\operatorname{Exec}_{\mathrm{intent}}(I,D)$ denotes the result implied by directly realizing intent $I$ on $D$.

\paragraph{Ambiguity.}
Without further clarification, $nl$ is considered ambiguous under $D$ when there exist at least two distinct plausible intents that can lead to different query results.
Formally, with $\mathcal{I}_D(nl)$ denoting the set of plausible user intents for question $nl$ under database $D$, there exist $I_i,I_j\in\mathcal{I}_D(nl)$ with $I_i\neq I_j$ and corresponding SQL queries $q_i$ and $q_j$ such that
\[
\begin{aligned}
    \operatorname{Exec}(q_i,D) &= \operatorname{Exec}_{\mathrm{intent}}(I_i,D), \\
    \operatorname{Exec}(q_j,D) &= \operatorname{Exec}_{\mathrm{intent}}(I_j,D), \\
    \operatorname{Exec}(q_i,D) &\neq \operatorname{Exec}(q_j,D).
\end{aligned}
\]

\paragraph{Setup and Problem Formulation.}

This work aims to resolve the ambiguity problem by obtaining additional specifications through user interaction. In our setup, an initial SQL prediction $q_0$ is generated at turn 0. When $q_0$ does not reflect the user's intended interpretation, the system obtains additional clarification from the user and progressively revises the prediction. We represent this process by an interaction trajectory
\begin{equation}
    \tau = \bigl(q_0,\mathcal{C}_1,\mathcal{F}_1,q_1,\ldots,\mathcal{C}_T,\mathcal{F}_T,q_T\bigr),
    \label{eq:interaction-trajectory}
\end{equation}
terminating after $T$ turns. At turn $t$, $\mathcal{C}_t=(\mathrm{cq}_{t,1},\ldots,\mathrm{cq}_{t,m_t})$ denotes the batch of CQs presented to the user, and $\mathcal{F}_t=(f_{t,1},\ldots,f_{t,m_t})$ denotes the corresponding user feedback. The system incorporates this feedback together with the preceding interaction history to produce an updated SQL prediction $q_t$. The interaction terminates when the current SQL is accepted or when the interaction budget is exhausted.  

Our goal is to obtain a terminal query whose execution matches the user's intended result while keeping the cost low. Let $\operatorname{Cost}_{\mathrm{tok}}(\tau)$ denote the total number of input and output tokens consumed by LLM calls along trajectory $\tau$. The objective is
\begin{equation}
    \min\; \operatorname{Cost}_{\mathrm{tok}}(\tau)
    \qquad
    \text{s.t.}
    \qquad
    \operatorname{Exec}(q_T,D) = \operatorname{Exec}_{\mathrm{intent}}(I^\star,D).
    \label{eq:interaction-objective}
\end{equation}

\subsection{Ambiguity Taxonomy}
\label{sec:taxonomy}

We follow the ambiguity taxonomy introduced by \sphinteract~\cite{sphinteract}, which categorizes Text-to-SQL ambiguity into four types:
\begin{itemize}
    \item \textsc{AmbColumn}: ambiguous mapping of an expression in the NL question to the schema, typically due to unclear column semantics or multiple columns with similar meanings.
    \item \textsc{AmbValue}: uncertainty in selecting the appropriate value for a predicate, often due to a mismatch between how a value is mentioned in the NL question and how it is stored in the database.
    \item \textsc{AmbOutput}: underspecified output requirements, such as which columns should be returned, how the result should be formatted, or in what order it should be presented.
    \item \textsc{AmbQuestion}: ambiguity in the semantics of the natural-language request itself, such as an unclear aggregation rule or an otherwise vague question.
\end{itemize}

\subsection{Interactive Disambiguation Protocol}
\label{sec:sphinteract}

\ours is based on the state-of-the-art interactive disambiguation protocol introduced by \sphinteract~\cite{sphinteract}, which we formally describe in this section.

\paragraph{\sphinteract's Interactive Protocol.}
Starting from an initial SQL prediction, the protocol repeatedly detects unresolved ambiguity, asks the user for clarification, and regenerates SQL. Let $D_{\mathrm{schema}}$ denote the database schema of $D$, and let $\tau_{<t}=(q_0,\mathcal{C}_1,\mathcal{F}_1,q_1,\ldots,\mathcal{C}_{t-1},$ $\mathcal{F}_{t-1},q_{t-1})$ denote the interaction history before turn $t$, following Eq.~\ref{eq:interaction-trajectory}.
At each turn, \sphinteract applies the Summarize--Review--Ask (SRA) procedure to produce one CQ:
\begin{equation}
    \mathrm{cq}_{t,1} = \operatorname{SRA}(nl,D_{\mathrm{schema}},\tau_{<t}).
    \label{eq:sphinteract-SRA}
\end{equation}
Within a single LLM call, SRA first summarizes the available context, including the NL question, schema, and history, then identifies the remaining ambiguity according to the four-type taxonomy, and generates a multiple-choice CQ. Since exactly one CQ is asked at each turn, $m_t=1$, $\mathcal{C}_t=(\mathrm{cq}_{t,1})$, and $\mathcal{F}_t=(f_{t,1})$.
After obtaining feedback $f_{t,1}$, the system regenerates the SQL as
\begin{equation}
    q_t = \operatorname{Regenerate}(nl,D_{\mathrm{schema}},\tau_{<t},\mathrm{cq}_{t,1},f_{t,1}).
    \label{eq:sphinteract-regenerate}
\end{equation}
The revised result is presented to the user again, and the process continues until the result is accepted or the interaction budget is exhausted.

Our work follows the overall \sphinteract's interactive protocol, including user clarification, feedback accumulation, SQL regeneration, and subsequent iterative rounds. Our contribution instead focuses on how relevant alternatives in the database are identified and supplied to the clarification process.


\begin{figure*}[t]
    \centering
    \includegraphics[width=\textwidth]{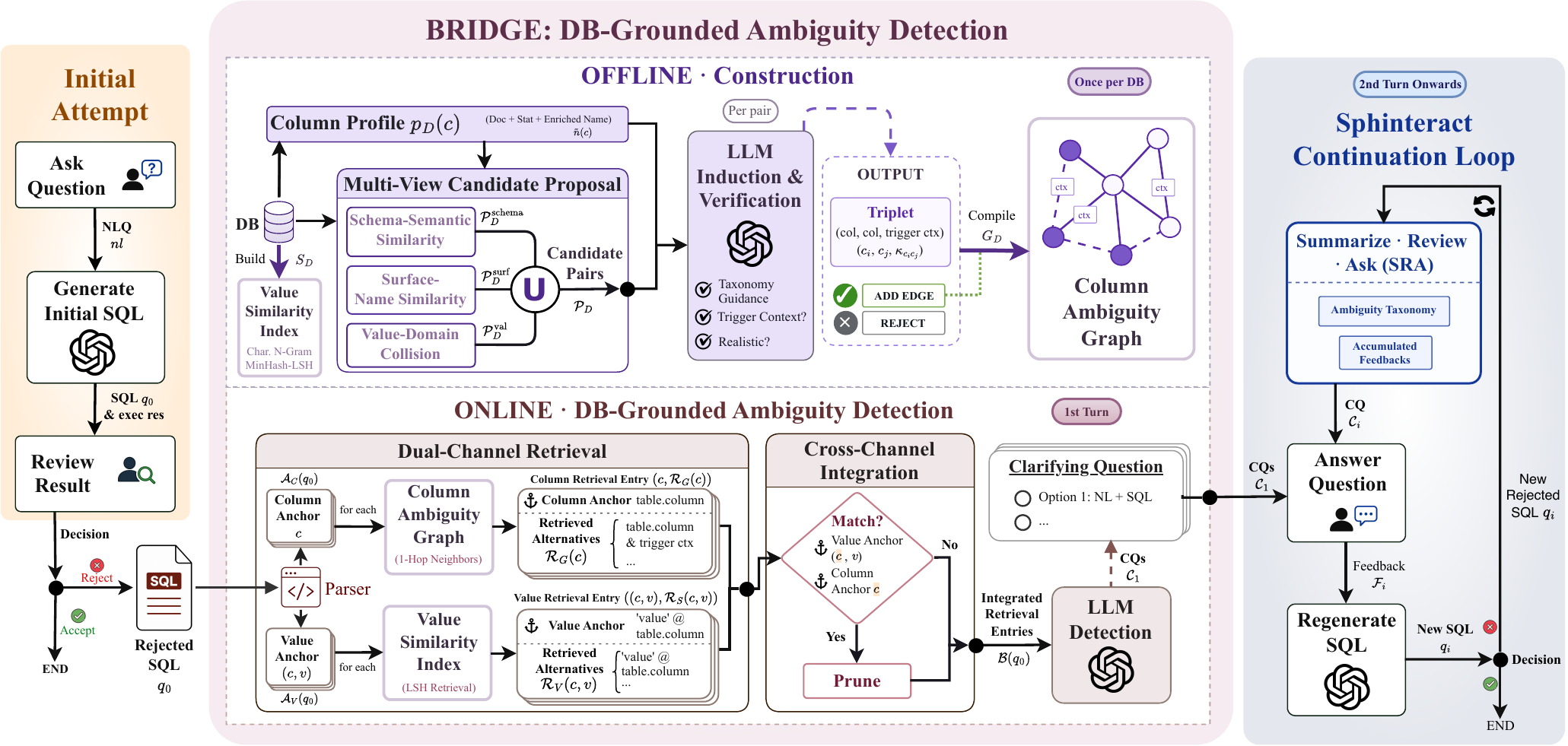}
    \caption{Overall framework of \ours and its integration with the existing iterative clarification protocol \cite{sphinteract}. }
    \label{fig:offline-online}
\end{figure*}

\section{Methodology}
\label{sec:methodology}

We first provide an overview of our method (Section~\ref{sec:method-overview}). We then describe the major component of the offline stage: Column Ambiguity Graph Construction (Section~\ref{sec:offline-graph}). Finally, we detail the online DB-Grounded Ambiguity Detection process (Section~\ref{sec:online-clarification}).

\subsection{Overview}
\label{sec:method-overview}
The key idea of \ours is to supply the LLM with offline-distilled reasoning cues about potential ambiguous relations among columns, which thus avoids solely relying on LLM parametric knowledge for ambiguity detection while reducing online LLM cost.
As illustrated in Figure~\ref{fig:offline-online}, \ours is integrated into the interactive disambiguation protocol described in Section~\ref{sec:sphinteract}. When the initial SQL prediction $q_0$ is rejected, \ours focuses on resolving column/value-grounded ambiguity at the first clarification turn. If the updated SQL is still rejected, subsequent turns continue with the \sphinteract protocol to address the remaining ambiguity types.

\paragraph{\ours Overview}
Formally, we first model the two reusable structures constructed offline as follows:
\begin{itemize}
    \item \textbf{Column Ambiguity Graph ($G_D$)}: An undirected graph whose nodes are database columns and whose edges indicate potentially ambiguous relations. For a connected column pair $c_i$ and $c_j$, the edge is annotated with $\kappa_{c_i c_j}$, which we call the \textbf{ambiguity trigger context}: a description of the condition under which the columns are potentially ambiguous. For example, in Figure~\ref{fig:running-example}, \texttt{schools.EdOpsName} and \texttt{schools.EILName} form such a pair, with the ambiguity trigger context being ``when requested for school category without specifying whether it refers to an education option or an instructional level.''
    \item \textbf{Value Similarity Index ($S_D$)}: A similarity index of textual values stored in the database. Following the value-retrieval design of CHESS~\cite{CHESS}, we normalize textual database values, represent them using character $n$-grams, and index them with MinHash-LSH for efficient approximate matching at query time. Each indexed value retains its source column, allowing retrieval to trace the corresponding database column.
\end{itemize}

With these two structures, the overall procedure of \ours is formalized as
\begin{align}
(G_D,S_D)
&=
\operatorname{BRIDGE}_{\mathrm{offline}}(D),
\label{eq:bridge-offline}\\
\mathcal{C}_1
&=
\operatorname{BRIDGE}_{\mathrm{online}}
\bigl(nl,q_0,D;\,G_D,S_D\bigr),
\label{eq:bridge-online}
\end{align}
where $D$ denotes the database, and $\mathcal{C}_1$ is the first-turn CQ batch in the interaction trajectory defined in Eq.~\ref{eq:interaction-trajectory}.

The \textbf{offline} stage creates $G_D$ and $S_D$ once for each database. In particular, \textbf{Column Ambiguity Graph Construction} performs two steps to construct $G_D$:
\begin{itemize}
    \item \textbf{Multi-View Candidate Proposal} identifies potentially competing column pairs using complementary signals from schema semantics, column names, and database values.
    \item \textbf{LLM Induction and Verification} then uses an LLM to derive the ambiguity trigger context and determine whether a pair should be added to $G_D$.
\end{itemize}

The \textbf{online} stage consists of \textbf{DB-Grounded Ambiguity Detection}, which produces $\mathcal{C}_1$ given the rejected $q_0$ in three steps:
\begin{itemize}
    \item \textbf{Dual-Channel Retrieval} uses columns and values mentioned in $q_0$ as anchors to retrieve column alternatives and similar value matches from the two offline-constructed artifacts, $G_D$ and $S_D$.
    \item \textbf{Cross-Channel Integration} follows designed rules to associate retrieved results from the two sources and filter unrelated alternatives.
    \item \textbf{LLM Detection} identifies ambiguity under the guidance of the remaining alternatives and produces CQs.
\end{itemize}

\paragraph{\sphinteract continuation loop.}
After obtaining user feedback $\mathcal{F}_1$ to $\mathcal{C}_1$, a new SQL query $q_1$ is regenerated following the procedure in Eq.~\ref{eq:sphinteract-regenerate}. If $q_1$ is still rejected, subsequent turns follow \sphinteract's protocol described in Section~\ref{sec:sphinteract}, but with the SRA procedure (Eq.~\ref{eq:sphinteract-SRA}) reviewing only \textsc{AmbOutput} and \textsc{AmbQuestion} to target ambiguity outside the column- and value-grounding issues.

\subsection{Column Ambiguity Graph Construction}
\label{sec:offline-graph}

The first artifact constructed offline is the Column Ambiguity Graph $G_D$, which captures column pairs that may become competing alternatives under an underspecified NL question. Constructing $G_D$ raises two challenges.

\textbf{Challenge 1: Which column pairs should be added to the graph?}
Exhaustively examining all column pairs is expensive and noisy, while not all relatedness, such as semantic similarity, necessarily implies plausible ambiguity.

\textbf{Challenge 2: How can the graph effectively and efficiently guide online ambiguity detection?}
A column pair alone does not explain why the columns may compete, whereas repeatedly providing detailed column profiles would incur substantial online cost.

We address Challenge~1 through coarse candidate proposal followed by LLM-based verification. For Challenge~2, we perform the DB-grounded reasoning offline and summarize it as a concise, reusable \emph{ambiguity trigger context} on each verified edge, describing the condition under which the two columns may become ambiguous.

\subsubsection{Schema and Data Profiling.}
\label{sec:schema-data-profiling}
We first perform a preliminary step to construct a profile $p_D(c)$ for each column $c$, which is used in later steps. The profile contains schema information, available documentation, sampled values, and lightweight value statistics, such as the number of unique values and the range of numerical values. Because raw column names can be abbreviated or opaque, we additionally use one LLM call per table to generate an \emph{enriched name} $\tilde n(c)$ for each column. For example, the enriched name of \texttt{schools.EILName} is ``Educational Instructional Level Name,'' while that of \texttt{schools.EdOpsName} is ``Educational Option Name.''

\subsubsection{Multi-View Candidate Proposal}
\label{sec:multi-view-proposal}

\begin{sloppypar}
We propose candidate column pairs from three views, corresponding to three types of database signals from which potentially ambiguous column relations may arise:
\begin{enumerate}
    \item[\textbf{(i)}] \textbf{Schema-semantic alternatives}, where columns describe the same or closely related concepts but differ in scope, granularity, representation, or semantic facet. For example, \texttt{gender.gender} and \texttt{gender.id} relate to the same concept but in different formats.
    \item[\textbf{(ii)}] \textbf{Surface-name collisions}, where column names appear lexically or semantically similar despite referring to different concepts. Examples include \texttt{results.position} vs \texttt{driverStandings.position}, and \texttt{legalities.format} vs \texttt{cards.layout}.
    \item[\textbf{(iii)}] \textbf{Value-domain collisions}, where columns contain overlapping values. For example, several languages appear in both \texttt{foreign\_data.language} and \texttt{set\_translations.} \\\texttt{language}.
\end{enumerate}
These signals are not mutually exclusive, as a column pair may exhibit more than one of them. For example, \texttt{foreign\_data.language} and \texttt{set\_translations.language} also exhibit a surface-name collision. 
\end{sloppypar}

Motivated by these signals, we design three corresponding proposal views: \textbf{Schema-Semantic Similarity}, \textbf{Surface-Name Similarity}, and \textbf{Value-Domain Collision}.

\paragraph{Schema-Semantic Similarity.}
This view targets columns whose meanings are closely related. Since the enriched name obtained in the preliminary step (Section \ref{sec:schema-data-profiling}) summarizes the meaning of a column in its schema context, embedding similarity can retrieve columns describing related concepts across different tables or naming conventions. We therefore embed the enriched name $\tilde n(c)$ of each column using an embedding model $f_{\mathrm{emb}}$. Similarity is measured by the inner product between normalized embeddings.
Formally, let $\mathrm{NN}_{k_{\mathrm{sem}}}(c)$ denote the top-$k_{\mathrm{sem}}$ nearest columns ranked by this similarity. We define
\begin{equation}
\begin{aligned}
\mathcal{P}^{\mathrm{schema}}_D
=
\bigl\{\{c_i,c_j\}\ \big|\ &
\bigl[c_j\in\mathrm{NN}_{k_{\mathrm{sem}}}(c_i)
\lor c_i\in\mathrm{NN}_{k_{\mathrm{sem}}}(c_j)\bigr],\\
&
\left\langle
f_{\mathrm{emb}}(\tilde n(c_i)),
f_{\mathrm{emb}}(\tilde n(c_j))
\right\rangle
\geq \delta_{\mathrm{sem}}
\bigr\},
\end{aligned}
\label{eq:schema-semantic-proposal}
\end{equation}
where $\delta_{\mathrm{sem}}$ is the similarity score threshold.

\paragraph{Surface-Name Similarity.}
The second view directly compares the original column names and targets potential ambiguity arising from surface-name collisions. 
We consider both lexical and semantic similarity. For lexical similarity, we use three complementary measures: \emph{word-level Jaccard similarity}, computed over lowercase word tokens split by whitespace, underscores, and case boundaries; \emph{character-trigram Dice similarity}; and \emph{edit similarity}. For semantic similarity, we use embedding similarity directly over the original column names to recover semantically related names.

A pair is proposed if any of these signals exceeds its corresponding threshold:
\begin{equation}
\begin{aligned}
\mathcal{P}^{\mathrm{surf}}_D
={}&
\bigl\{\{c_i,c_j\}\ \big|\
\operatorname{Jaccard}(c_i,c_j)\geq\delta_{\mathrm{jac}}
\ \lor\
\operatorname{Dice}_{3}(c_i,c_j)\geq\delta_{\mathrm{dice}}\\
&
\operatorname{EditSim}(c_i,c_j)\geq\delta_{\mathrm{edit}}
\ \lor\
\operatorname{EmbSim}(c_i,c_j)\geq\delta_{\mathrm{name}}
\bigr\}.
\end{aligned}
\label{eq:surface-name-proposal}
\end{equation}
Here, $\operatorname{Jaccard}$, $\operatorname{Dice}_3$, and $\operatorname{EditSim}$ denote the three lexical similarity measures, while $\operatorname{EmbSim}$ denotes embedding similarity over the original column names. The corresponding $\delta$ terms denote their similarity thresholds.

\paragraph{Value-Domain Collision.}
The third view targets columns whose stored textual values overlap. We apply this proposal view only to columns containing textual values, further excluding Boolean, date-like, and numeric columns whose value overlap is unlikely to indicate useful ambiguity. Formally, let $\mathcal{V}(c)$ denote the set of distinct non-null values in column $c$. We propose a pair when the overlap, normalized by the size of the smaller value set, exceeds threshold $\delta_{\mathrm{val}}$:
\begin{equation}
\mathcal{P}^{\mathrm{val}}_D
=
\left\{
\{c_i,c_j\}
\;\middle|\;
\frac{
|\mathcal{V}(c_i)\cap\mathcal{V}(c_j)|
}{
\min\{|\mathcal{V}(c_i)|,|\mathcal{V}(c_j)|\}
}
\geq \delta_{\mathrm{val}}
\right\}.
\label{eq:value-domain-proposal}
\end{equation}

\paragraph{Candidate Aggregation.}
The final candidate set is the union of the three proposal views:
\begin{equation}
\mathcal{P}_D
=
\mathcal{P}^{\mathrm{schema}}_D
\cup
\mathcal{P}^{\mathrm{surf}}_D
\cup
\mathcal{P}^{\mathrm{val}}_D.
\label{eq:multi-view-proposal}
\end{equation}
A candidate pair may be proposed by multiple views. For example, \texttt{schools.EdOpsName} and \texttt{schools.EdOpsCode} is proposed through both schema-semantic and surface-name similarity, whereas pairs involving columns \texttt{schools.EdOpsName}, \texttt{schools.EILName}, and \texttt{frpm.School Type} are proposed through schema-semantic similarity. 

\subsubsection{LLM Induction and Verification}
\label{sec:llm-induction-verification}

As discussed above, we use the LLM for two purposes: to induce an ambiguity trigger context, and to verify whether the induced condition is sufficiently realistic to retain the pair in $G_D$. Since these two tasks naturally form a reasoning process, we instruct the LLM to perform them sequentially within a single call.

Formally, for each candidate pair $\{c_i,c_j\}\in\mathcal{P}_D$, the LLM receives the corresponding column profiles and produces an ambiguity trigger context together with a verification decision:
\begin{equation}
\begin{aligned}
(\kappa_{c_i c_j},d_{ij})
&=
\operatorname{LLM}
\bigl(p_D(c_i),p_D(c_j)\bigr),\\
d_{ij}
&\in
\{\textsc{AddEdge},\textsc{Reject}\}.
\end{aligned}
\label{eq:pair-verifier}
\end{equation}
Here, $p_D(c_i)$ and $p_D(c_j)$ are the profiles of columns $c_i$ and $c_j$, containing information such as schema descriptions, representative values, and value statistics, which help the LLM understand the columns beyond their names alone. $\kappa_{c_i c_j}$ denotes the induced ambiguity trigger context. $d_{ij}$ denotes the verification decision.
The prompt instructs the LLM to first consider what a user could realistically request and what missing specification would make both columns plausible interpretations. It summarizes this condition as $\kappa_{c_i c_j}$, the ambiguity trigger context introduced in Section~\ref{sec:method-overview}. It then determines whether the induced situation represents a realistic and reusable source of ambiguity rather than an artificial corner case.
To guide the reasoning, the prompt includes these reasoning lenses together with typical examples, describing common ways in which two columns may become ambiguous:
\begin{itemize}
    \item \emph{Shared referent}: the columns refer to the same underlying entity, attribute, or semantic object but use different identifiers, names, codes, representations, or levels of detail. A typical case is multiple representations of the same concept, such as an ID/code and a name.
    
    \item \emph{Conceptual facets}: the columns represent distinct facets, dimensions, components, or stages of the same higher-level concept or analytical frame. For example, \texttt{surname} and \texttt{forename}; or \texttt{campaigns.views} and \texttt{campaigns.clicks} represent different facets of performance.
    
    \item \emph{Contextual collision}: the columns have similar names or semantic roles but belong to different entity, table, event, row-grain, or task contexts. For example, \texttt{products.type} and \texttt{orders.category} have similar semantic roles but apply to different targets.
\end{itemize}
These lenses provide reasoning guidance rather than an exhaustive taxonomy, and the LLM is not required to assign every candidate pair to one of them.

For the running example (Figure \ref{fig:running-example}), reasoning about the pair \texttt{schools.EdOpsName} and \texttt{schools.EdOpsCode} is naturally guided by the \emph{shared referent} lens because the two columns provide the name and code of the same concept, an education option. In contrast, reasoning about \texttt{frpm.} \texttt{School Type} and \texttt{schools.EILName} can be guided by \emph{conceptual facets}, as both relate to the broader concept of school category but in different categorization ways.

For every LLM-verified pair with $d_{ij}=\textsc{AddEdge}$, we add an undirected edge between $c_i$ and $c_j$ to $G_D$ and attach the induced trigger context $\kappa_{c_i c_j}$ as its annotation.

\subsection{DB-Grounded Ambiguity Detection}
\label{sec:online-clarification}
Given the rejected $q_0$, the online stage uses the columns and values mentioned in $q_0$ to retrieve possible alternatives from the two offline-constructed artifacts, $G_D$ and $S_D$. It then integrates the retrieval results and asks the LLM to produce the CQ batch $\mathcal{C}_1$.

\subsubsection{Dual-Channel Retrieval}
\label{sec:dual-channel-retrieval}

We first parse the database columns and values mentioned in $q_0$, which we refer to as \textbf{column anchors} and \textbf{value anchors}, respectively:
\begin{equation}
\begin{aligned}
\mathcal{A}_C(q_0)
&=
\{c \mid c \text{ is a column referenced in } q_0\},\\
\mathcal{A}_V(q_0)
&=
\bigl\{(c,v) \ \big|\ 
v \text{ is a value referenced in } q_0,\\
&\hspace{3.3em}
v \text{ is associated with column } c
\bigr\}.
\end{aligned}
\label{eq:online-anchor-sets}
\end{equation}
We use SQLGlot~\cite{sqlglot} to parse $q_0$ and resolve column references to the original database schema.

\begin{sloppypar}
The rejected SQL provides a useful starting point for retrieval because it represents the initial grounding of the NL question to the database. Rather than searching for ambiguity across the entire database, \ours retrieves alternatives around the database elements already selected in $q_0$, naturally focusing retrieval on elements relevant to the current question. For example, in Figure~\ref{fig:running-example}, the initial SQL interprets ``educational level name'' using \texttt{schools.EdOpsName} and filters ``Breakfast Provision 2'' using \texttt{frpm.NSLP Provision Status}, which become a column and value anchor, respectively.
\end{sloppypar}

The anchors are then used by two complementary retrieval channels: \textbf{Graph-Based Retrieval} and \textbf{Value-Index-Based Retrieval}.

\paragraph{Graph-Based Retrieval.}
For each column anchor $c\in\mathcal{A}_C(q_0)$, we retrieve all \emph{one-hop neighbors} of $c$ in the Column Ambiguity Graph $G_D$:
\begin{equation}
\mathcal{R}_{G}(c)
=
\left\{
(c',\kappa_{cc'})
\mid
c' \text{ is adjacent to } c \text{ in } G_D
\right\}.
\label{eq:graph-retrieval}
\end{equation}
\begin{sloppypar}
Each retrieved $c'$ is considered as a potential \emph{column alternative} with its ambiguity trigger context $\kappa_{cc'}$ describing the reason. 
In the running example (Figure \ref{fig:running-example}), retrieved alternatives for the column anchor \texttt{schools.EdOpsName} include \texttt{schools.EILName}, \texttt{frpm.} \texttt{School Type}, and \texttt{schools.EdOpsCode}, together with the ambiguity trigger context describing under what condition each column may compete with \texttt{schools.EdOpsName}. 
\end{sloppypar}

We consider one-hop retrieval sufficient because an edge in $G_D$ already represents a directly verified potential ambiguity relation between its two endpoint columns.

\paragraph{Value-Index-Based Retrieval.}
For each value anchor $(c,v)\in$ \\ $\mathcal{A}_V(q_0)$, we perform LSH similarity search to find potential \emph{value alternatives} in the Value Similarity Index $S_D$ using $v$:
\begin{equation}
\begin{aligned}
\mathcal{R}_{S}(c,v)
=
\bigl\{(c',v') \ \big|\ &
v' \text{ is retrieved from } S_D \text{ using } v,\\
&
\text{and belongs to column } c'
\bigr\}.
\end{aligned}
\label{eq:value-retrieval}
\end{equation}

\begin{sloppypar}
For the example in Figure~\ref{fig:running-example}, querying \texttt{'Breakfast Provision 2'} in the value index retrieves its occurrence under \texttt{frpm.NSLP Provision Status}, and several similar values under the same column: \texttt{'Breakfast Provision 1'} and \texttt{'Breakfast Provision'}.
\end{sloppypar}

We introduce this channel for two purposes. First, a value alternative may provide a complementary signal for identifying an alternative column that is not connected to the current column in $G_D$. Second, it can expose \textsc{AmbValue} when different plausible representations are stored in the database.

\subsubsection{Cross-Channel Integration}
\label{sec:cross-channel-integration}
\begin{sloppypar}
Column and value anchors can describe the same local interpretation decision, so treating their retrieved alternatives independently may retain unnecessary alternatives and lead to redundant or irrelevant CQs.
The running example in Figure~\ref{fig:running-example} illustrates such a situation. The column anchor \texttt{frpm.NSLP Provision Status} and the value anchor $(\texttt{frpm.NSLP Provision Status}, \texttt{'Breakfast Provision 2'})$ \\ correspond to the same filtering decision in $q_0$: how ``schools with Breakfast Provision 2'' should be interpreted. Graph-based retrieval identifies \texttt{schools.StatusType} as a potential alternative, but value-index-based retrieval does not retrieve \texttt{'Breakfast Provision 2'} under this column, suggesting that it is not a valid alternative under the current NL question and can be excluded.
\end{sloppypar}
We observe that such overlap between the two channels can be identified through the column to which their anchors are grounded. In the example above, both anchors are grounded to \texttt{frpm.NSLP Provision Status}. Cross-Channel Integration therefore associates retrieved alternatives from the two channels according to the shared grounded column of their anchors and prunes incompatible column alternatives.

Algorithm~\ref{alg:cross-channel-integration} formalizes this procedure. We refer to $(c,\mathcal{R}_{G}(c))$ and $((c,v),\mathcal{R}_{S}(c,v))$ as \textbf{column retrieval entries} and \textbf{value retrieval entries}, respectively. The integrated set of retained retrieval entries is denoted by $\mathcal{B}(q_0)$, which will be used in the subsequent LLM Detection (Section~\ref{sec:llm-detection}).
Notably, each distinct anchor contributes at most one retrieval entry, regardless of how many times it occurs in $q_0$, to avoid redundancy in the prompt.
Entries grounded to the same column appear continuously in $\mathcal{B}(q_0)$ (Lines~3--4 and 9--13 in Algorithm~\ref{alg:cross-channel-integration}), facilitating more focused detection reasoning.

\begin{algorithm}[t]
\caption{Cross-Channel Integration}
\label{alg:cross-channel-integration}
\begin{algorithmic}[1]
\Statex \textbf{Input:} Rejected SQL $q_0$; retrieved alternatives
$\{\mathcal{R}_{G}(c)\mid c\in\mathcal{A}_C(q_0)\}$ and
$\{\mathcal{R}_{S}(c,v)\mid(c,v)\in\mathcal{A}_V(q_0)\}$
\Statex \textbf{Output:} Integrated retrieval entries $\mathcal{B}(q_0)$
\State $\mathcal{B}(q_0) \gets \emptyset$
\ForAll{$c \in \mathcal{A}_C(q_0)$}
    \State $\textit{columnAlts} \gets \mathcal{R}_{G}(c)$
    \State $\textit{valueEntries} \gets
    \{((c,v),\mathcal{R}_{S}(c,v)) \mid (c,v)\in\mathcal{A}_V(q_0)\}$
    \State $\textit{supportedCols} \gets
    \{c' \mid \exists\,((c,v),R)\in\textit{valueEntries},\;
    \exists\,v' \text{ s.t. } (c',v')\in R\}$
    \If{every occurrence of $c$ in $q_0$ is associated with a value anchor occurrence}
        \State $\textit{columnAlts} \gets
        \{(c',\kappa_{cc'})\in\textit{columnAlts}
        \mid c'\in\textit{supportedCols}\}$
    \EndIf
    \If{$|\textit{columnAlts}|\geq 1$}
        \State $\mathcal{B}(q_0) \gets
        \mathcal{B}(q_0)\cup\{(c,\textit{columnAlts})\}$
    \EndIf
    \ForAll{$((c,v),R)\in\textit{valueEntries}$ such that $|R|>1$}
        \State $\mathcal{B}(q_0) \gets
        \mathcal{B}(q_0)\cup\{((c,v),R)\}$
    \EndFor
\EndFor
\State \Return $\mathcal{B}(q_0)$
\end{algorithmic}
\end{algorithm}

\begin{sloppypar}
For each column anchor $c$, the algorithm first collects all value anchors grounded to the same column and identifies the columns supported by their retrieved value alternatives. Pruning is applied only when every occurrence of $c$ in $q_0$ is associated with a value anchor occurrence. This condition is necessary because different occurrences of the same column may participate in different interpretation decisions.
Consider a modified version of the running example: ``What is the educational level name for the schools with Breakfast Provision 2 in county code 37? Also output the name of the school and status.'' Suppose the initial SQL is:
\texttt{SELECT s.EdOpsName, s.School, f.'NSLP Provision Status' FROM frpm AS f JOIN schools AS s ON f.CDSCode = s.CDSCode WHERE f.'NSLP Provision Status' = 'Breakfast Provision 2' AND f.'County Code' = 37}.
The column \texttt{frpm.NSLP Provision Status} occurs twice, but only the predicate occurrence is associated with a value anchor. The value retrieval informs the filtering decision but provides no information about whether the requested output ``status'' should use \texttt{frpm.NSLP Provision Status} or \texttt{schools.StatusType}. 
Since this column only corresponds to one retrieval entry in $\mathcal{B}(q_0)$, its retrieved alternatives must preserve all that is relevant to both occurrences.
Pruning should therefore not be applied in this case.
\end{sloppypar}

Finally, only retrieval entries that contain at least one alternative beyond the original anchor are retained in $\mathcal{B}(q_0)$; otherwise, the retrieval result does not suggest any competing alternative for the anchor and can thus be safely discarded. Since $\mathcal{R}_{G}(c)$ contains only neighboring columns, a column retrieval entry is retained when $|\mathcal{R}_{G}(c)|\geq 1$. In contrast, $\mathcal{R}_{S}(c,v)$ may also return the anchor's own occurrence, so the condition is $|\mathcal{R}_{S}(c,v)|>1$. The resulting $\mathcal{B}(q_0)$ is supplied to the subsequent LLM detection step.

\subsubsection{LLM Detection}
\label{sec:llm-detection}
The final step uses an LLM to examine the retained retrieval entries in $\mathcal{B}(q_0)$ and produce the CQ batch $\mathcal{C}_1$. Formally,
\[
\mathcal{C}_1
=
\operatorname{LLM}
\bigl(nl,q_0,D'_{\mathrm{schema}},\mathcal{B}(q_0)\bigr).
\]
Here, $nl$ and $q_0$ denote the NL question and the rejected SQL prediction, respectively. $D'_{\mathrm{schema}}$ denotes the relevant schema subset, containing the columns referenced in $q_0$, all columns appearing in $\mathcal{B}(q_0)$, their associated table information, and the database-defined foreign-key relations among the involved tables.

\begin{sloppypar}
The LLM may further exclude irrelevant alternatives by reasoning over the ambiguity trigger contexts and specific values, which we refer to as \emph{LLM contextualization}.
It then formulates CQs for the ambiguity issues it identifies. Each generated CQ consists of a CQ stem and a set of candidate interpretations. Each candidate contains a short NL description and a minimal SQL snippet, which are concatenated to form the final clarification option.
In the example in Figure~\ref{fig:running-example}, the LLM excludes \texttt{schools.EdOpsCode} and produces a CQ about ``educational level name'' with options spanning the anchor \texttt{schools.EdOpsName} and two school-category columns, \texttt{frpm.School Type} and \texttt{schools.EILName}.
\end{sloppypar}


\section{Experiments}
\label{sec:experiments}

\subsection{Settings}
\label{sec:experiment-settings}

\subsubsection{Dataset.}
\label{sec:datasets}
We evaluate on BIRD-dev~\cite{BIRD}, a cross-domain Text-to-SQL benchmark containing 1,534 natural-language questions over 11 relational databases. Since substantial ground-truth inaccuracies have been reported~\cite{SQLDriller,benchmarkBroken}, we use the corrected version provided by our previous work~\cite{hisql}, except for cases whose correction expands the set of gold SQLs due to ambiguity issues, for which we retain the original single gold SQL. 
Additionally, BIRD provides \textit{evidence} for each case that supplies additional specifications for each question. Since much of the evidence provides specifications related to potential ambiguity, we exclude it from our main experiments to create a more challenging setting that stress-tests the evaluated methods. For completeness, we also report results with the provided evidence in Section~\ref{sec:with-bird-evidence}.

\subsubsection{Evaluation Metrics.}
Following \cite{hisql}, we use column-order-insensitive execution accuracy, denoted by $\mathrm{EX}_{\mathrm{col}}$. Let $\hat q_i$, $q_i^\star$, and $D_i$ denote the predicted SQL, gold SQL, and database for case $i$, respectively. A prediction is considered correct if there exists a permutation of its output columns under which its execution result matches that of the reference SQL:
\begin{equation}
\mathrm{EX}_{\mathrm{col}}
=
\frac{1}{N}
\sum_{i=1}^{N}
\mathbb{I}
\left[
\exists\,\pi:
\pi\!\left(\operatorname{Exec}(\hat q_i,D_i)\right)
=
\operatorname{Exec}(q_i^\star,D_i)
\right],
\label{eq:ex-col}
\end{equation}
where $N$ is the number of evaluation cases, $\mathbb{I}[\cdot]$ is the indicator function, and $\pi$ denotes a permutation applied to the output columns of the execution result.

Following the cost formulation in Section~\ref{sec:problem-formulation}, our primary efficiency metric is \emph{LLM cost}, defined as the average cumulative input and output tokens consumed by system-side LLM calls per case. We exclude calls used solely for user simulation, as we follow \sphinteract~\cite{sphinteract} to use an LLM as a user stimulator to answer CQs, and these calls represent the external user rather than the evaluated system. In some analysis experiments, the number of 
CQs per case and the number of choices per CQ are used as metrics as well.

\subsubsection{Baselines}
\label{sec:baselines}

We compare \ours with four recent interactive Text-to-SQL disambiguation systems.

\begin{itemize}
    \item \textbf{\textsc{Sphinteract}}~\cite{sphinteract}, as described in Section~\ref{sec:preliminaries}, reviews the interaction history and generates a CQ guided by the taxonomy (Section~\ref{sec:taxonomy}), then regenerates the SQL using the received feedback. This process continues until the SQL is accepted or the interaction budget is exhausted.

    \item \textbf{\textsc{AtomSQL}}~\cite{AtomSQL} generates candidate SQL queries from multiple models, decomposes each query into clauses, and uses model disagreement to expose uncertainty. The user makes clause-level selections to identify the highest-scoring SQL.

    \item \textbf{\textsc{CLEAR}}~\cite{clear} uses an LLM-based detector and a rule-based corrector to construct candidate mappings. Each mapping represents choices of column usage for an ambiguous expression in the NL question. After obtaining the user-selected mapping, it rewrites the NL question and uses it for SQL prediction. Following its evaluation setting, we simulate user selection deterministically by choosing the candidate mapping according to the columns used in the gold SQL.

    \item \textbf{\textsc{AmbiSQL}}~\cite{AmbiSQL} uses a fine-grained ambiguity taxonomy to detect ambiguities, maintains candidate interpretations in a preference tree, and incorporates the selected interpretation into the clarified NL question before invoking the downstream parser.
\end{itemize}

\begin{figure*}[th]
    \centering
    \includegraphics[width=0.9\linewidth]{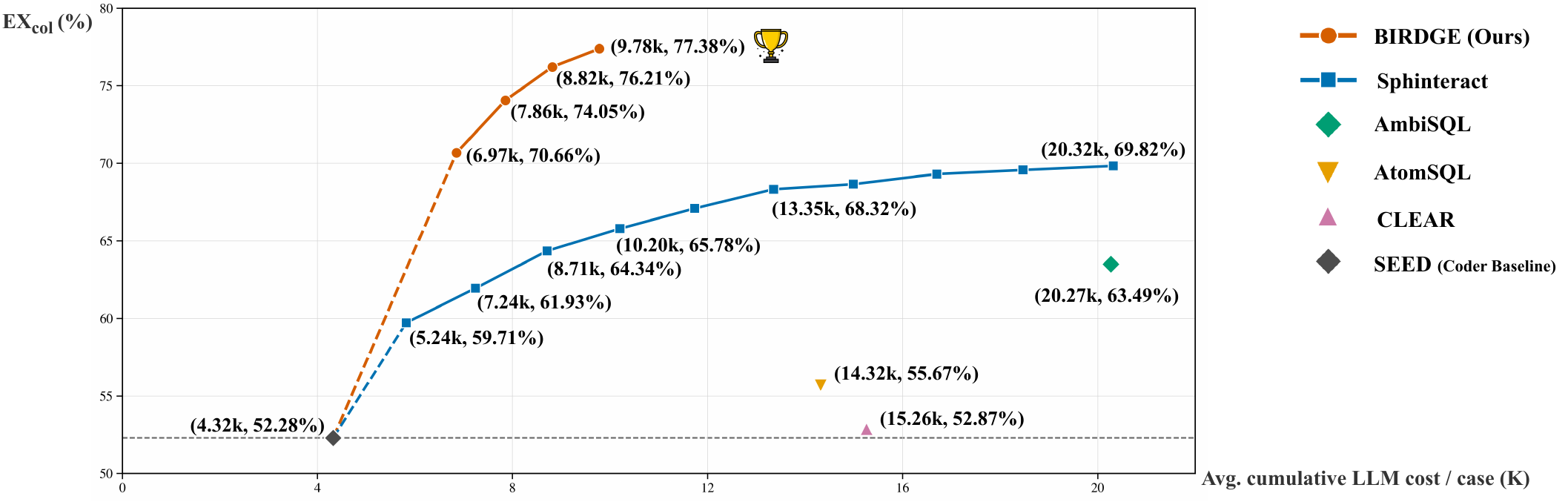}
    \caption{
    End-to-end accuracy--cost trade-off on the complete BIRD-dev without the benchmark-native evidence. The average cumulative LLM cost per question excludes that for user simulation. \seed denotes the common seed result used by \ours and \textsc{Sphinteract}. For \ours, the points correspond to T1--T4; for \textsc{Sphinteract}, we show representative intermediate turns (T1--T4 and T6) and the final turn result. Final statistics are shown for \textsc{AtomSQL}, \textsc{CLEAR} and \textsc{AmbiSQL}. 
    }
    \label{fig:overall-accuracy-cost}
\end{figure*}

\subsubsection{Implementation Details}
\label{sec:implement-details}

\paragraph{Model Configuration.}
We use \offlinemodel~\cite{Qwen37Plus} for offline graph construction; \onlinemodel~\cite{qwen35} for online ambiguity handling, user simulation, and SQL regeneration; and \embeddingmodel~\cite{Qwen3Embedding} for embedding-based similarity and demonstration retrieval, unless otherwise specified. All LLM calls are made with reasoning disabled and temperature set to $0$.

\paragraph{Base Text-to-SQL Parser.}
Following \sphinteract~\cite{sphinteract}, we use a simple LLM-based Text-to-SQL parser to provide the initial SQL prediction but provide additional information than the schema only. Given an NL question, we prompt \seedmodel with the complete database schema with the benchmark-provided descriptions, sampled column values, and lightweight database statistics such as the numbers of unique and null values to generate one SQL query using greedy decoding. We use the same fixed parser to produce the initial seed SQL for \ours and \textsc{Sphinteract}, and as the downstream parser for \textsc{CLEAR} and \textsc{AmbiSQL} after their disambiguation stages. \textsc{AtomSQL} instead requires a multi-model candidate pool. We therefore generate its candidate SQLs using \seedmodel, \onlinemodel, and OmniSQL-32B~\cite{OmniSQL}, representing coding-specialized, general-purpose, and SQL-specialized generators, respectively.

\paragraph{User Simulation.}
Following \sphinteract~\cite{sphinteract}, we use an LLM-based user simulator that has access to the gold SQL and selects among the clarification choices. To keep the ability of the user simulator consistent across compared methods and settings, we consistently use \onlinemodel throughout all experiments. 

\paragraph{Parameter Settings.}
\ours uses the following parameters to balance the coverage of potentially relevant alternatives against the introduction of irrelevant noise and excessive context for subsequent LLM reasoning. 
Offline, the schema-semantic channel uses $k_{\mathrm{sem}}=10$ and $\delta_{\mathrm{sem}}=0.55$. The surface-name channel proposes a pair when word-level Jaccard similarity is at least $0.30$, character-trigram Dice similarity is at least $0.60$, normalized edit similarity is at least $0.70$, or name-embedding cosine similarity is at least $0.60$. The value-domain proposal uses an overlap coefficient threshold of $0.50$, and pair-verification profiles include up to five randomly sampled values per column together with value-overlap examples when applicable. 
Online, the value similarity index uses character trigrams ($n=3$) and returns at most $20$ matches per value anchor to limit retrieval noise and prompt size. Following \sphinteract~\cite{sphinteract}, SQL regeneration retrieves three in-context demonstrations from its user-study question bank using \embeddingmodel.

\subsection{Overall Performance}
\label{sec:overall-performance}

We evaluate the end-to-end effectiveness and efficiency of \ours against the four baselines: \textsc{CLEAR}~\cite{clear}, \textsc{AtomSQL}~\cite{AtomSQL}, \textsc{AmbiSQL}~\cite{AmbiSQL}, and \sphinteract~\cite{sphinteract}. We measure effectiveness using $\mathrm{EX}_{\mathrm{col}}$ and efficiency using the average cumulative LLM cost per case. The reported cost is averaged over all 1,534 BIRD-dev cases.

The results are shown in Figure~\ref{fig:overall-accuracy-cost}. Overall, \ours provides the strongest accuracy--cost trade-off among all compared methods. We compare \ours at final turn with the final results of the baselines, except for \sphinteract, for which we use T4 where its LLM cost is closest to the final cost of \ours. \ours achieves 77.38\% accuracy with a cost of 9.78K tokens per case, corresponding to accuracy improvements ranging from 11.60\% to 24.51\% while reducing LLM cost by up to 51.75\%.
\begin{sloppypar}
A closer comparison with the strongest baseline, \sphinteract, further demonstrates the effectiveness of our database-grounded ambiguity detection. At T1, where \ours specifically targets column- and value-grounding ambiguity, it already achieves 70.66\% accuracy at 6.97K tokens, exceeding even the 69.82\% reached by \sphinteract after ten turns while reducing the average LLM cost by 65.70\%.
The improvement comes from more targeted CQs: across all four turns, \ours achieves 11.60\% higher accuracy with a comparable clarification burden, asking 1.97 CQs per case on average compared with 1.64 for \sphinteract by T4.
Additionally, \sphinteract shows diminishing gains in later turns, improving by only 1.50\% from T6 (68.32\%) to T10 (69.82\%). We observe that later turns often produce CQs similar to those already asked.
Figure~\ref{fig:running-example} illustrates these limitations: \sphinteract explores nearly all plausible aspects of the request and even revisits some of them, yet fails to expose the critical issue involving the intended column \texttt{schools.EILName}.
Together, these observations reflect the difficulty of discovering critical alternatives through repeated LLM reasoning alone, whereas the competing column relations retrieved by \ours directly guide the LLM toward the critical distinction.
\end{sloppypar}
The remaining baselines are substantially less effective. \textsc{CLEAR}'s low accuracy (52.87\%) is due to insufficient ambiguity detection, as it detects and resolves ambiguity in only 84 of the 1,534 cases, leading to the corrections of just 10 cases.
\textsc{AtomSQL}'s selection-based framework has two inherent limitations: (1) candidates may be incorrect in different clauses, and clause-level reranking cannot combine their individually correct parts into an accurate SQL; and (2) performance is bounded by candidate quality. We observe that the upper bound of its candidate pool---counting a case as correct whenever any candidate is correct---reaches only 64.08\%, while a substantial gap of 8.41\% remains between this upper bound and the final accuracy (55.67\%).
\textsc{AmbiSQL} performs better (63.49\%) but requires the highest cost (20.27K tokens per case) among the compared interactive methods.

\subsection{Further Analysis}

\subsubsection{Column-Linking Analysis}
\label{sec:scoped-recall}

Overall execution accuracy measures whether the final SQL is correct, but does not directly reveal whether the interaction resolves the column-grounding errors targeted by \ours. We therefore conduct a column-linking analysis to examine how effectively \ours corrects column usage by comparing it with the strongest baseline, \textsc{Sphinteract}~\cite{sphinteract}.

\textit{Metrics.}
For case $i$, let $G_i$ and $T_i^{(t)}$ denote the sets of original database columns referenced by the gold SQL and the predicted SQL at turn $t$, respectively, with $T_i^{(0)}$ denoting the seed prediction. We quantify column-linking quality using micro-averaged \emph{precision}, \emph{recall}, and \emph{F1}:
\[
\mathrm{TP}_i^{(t)}=|G_i\cap T_i^{(t)}|,\qquad
\mathrm{FP}_i^{(t)}=|T_i^{(t)}\setminus G_i|,\qquad
\mathrm{FN}_i^{(t)}=|G_i\setminus T_i^{(t)}|.
\]
\begin{equation}
\begin{aligned}
R_\mu^{(t)} &=
\frac{\sum_i \mathrm{TP}_i^{(t)}}
{\sum_i(\mathrm{TP}_i^{(t)}+\mathrm{FN}_i^{(t)})},
\qquad
P_\mu^{(t)} =
\frac{\sum_i \mathrm{TP}_i^{(t)}}
{\sum_i(\mathrm{TP}_i^{(t)}+\mathrm{FP}_i^{(t)})},\\
F1_\mu^{(t)} &=
\frac{2\sum_i \mathrm{TP}_i^{(t)}}
{2\sum_i \mathrm{TP}_i^{(t)}+\sum_i \mathrm{FP}_i^{(t)}+\sum_i \mathrm{FN}_i^{(t)}}.
\end{aligned}
\label{eq:schema-linking-metrics}
\end{equation}
Here, $\mathrm{TP}_i^{(t)}$, $\mathrm{FP}_i^{(t)}$, and $\mathrm{FN}_i^{(t)}$ denote the true positives, false positives, and false negatives, respectively, of the predicted column set relative to the gold column set.

\begin{figure}[h]
    \centering
    \includegraphics[width=0.5\textwidth]{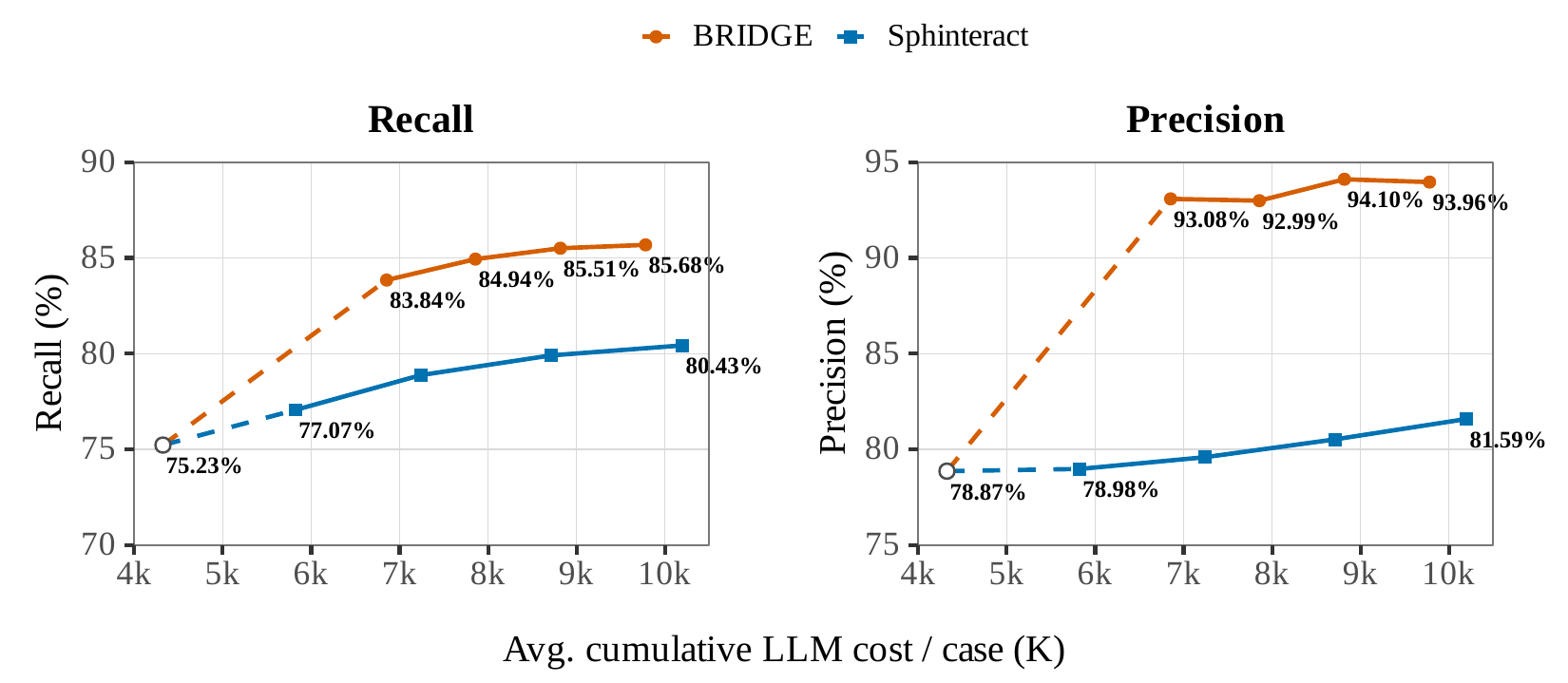}

    \vspace{2pt}
    \textbf{F1(\%)}

    \vspace{2pt}
    \begin{tabular}{@{}l c c c c c@{}}
        \toprule
        \textbf{Method}
        & \seed
        & $\mathbf{T1}$
        & $\mathbf{T2}$
        & $\mathbf{T3}$
        & $\mathbf{T4}$ \\
        \midrule
        \ours
        & 77.01
        & \textbf{88.22}
        & \textbf{88.78}
        & \textbf{89.60}
        & \textbf{89.63} \\
        \textsc{Sphinteract}
        & 77.01
        & 78.01
        & 79.24
        & 80.22
        & 81.01 \\
        \bottomrule
    \end{tabular}

    \caption{Schema-linking recall, precision, and F1.}
    \label{fig:schema-linking-analysis}
\end{figure}

Figure~\ref{fig:schema-linking-analysis} plots recall and precision against cumulative LLM cost and reports F1 for the shared seed and the first four turns of both methods.
\ours produces an immediate improvement in both recall and precision at T1: recall rises from 75.23\% to 83.84\%, while precision increases from 78.87\% to 93.08\%, yielding an F1 improvement from 77.01\% to 88.22\%. Subsequent turns provide smaller refinements, reaching 85.68\% recall, 93.96\% precision, and 89.63\% F1 at T4. In contrast, \textsc{Sphinteract} improves more gradually, reaching only 80.43\% recall, 81.59\% precision, and 81.01\% F1 at T4. The simultaneous gains in recall and precision show that \ours does not improve coverage merely by introducing more columns: it both recovers missing gold columns and suppresses spurious ones.

To further connect column-linking quality with execution accuracy, we examine failed predictions with incorrect column usage. We regard a failed seed as containing a \emph{potential column-related error} when $G_i \neq T_i^{(0)}$. Among the 732 incorrect seed predictions, 516 (70.49\%) satisfy this condition. After four turns, 64.53\% of these cases remain incorrect in both execution and column usage under \textsc{Sphinteract}, compared with only 35.85\% under \ours. 

All these results support the effectiveness of database-grounded clarification in resolving column-related errors.

\subsubsection{Effectiveness of the Ambiguity Graph}
\label{sec:graph-ablation}

We examine whether preconstructing the ambiguity graph provides useful information for ambiguity detection. We compare \ours with two variants. \textbf{w/o Graph} disables \emph{Graph-Based Retrieval} and instead provides the full schema directly to \emph{LLM Detection} for reasoning. \textbf{rp/w Query-Time} replaces \emph{Graph-Based Retrieval} with query-time retrieval, using one LLM call per column anchor to search the full schema for potentially competing columns and their corresponding ambiguity reasons. We report T1 LLM cost and the number of CQ per case, together with turn-wise $\mathrm{EX}_{\mathrm{col}}$.

\begin{table}[h]
    \centering
    \caption{Effectiveness of the ambiguity graph.}
    \label{tab:graph-effectiveness}
    \small
    \setlength{\tabcolsep}{4pt}
    \begin{tabular}{@{}l c c c c c c@{}}
        \toprule
        \textbf{Setting}
        & \multicolumn{2}{c}{\textbf{T1 Avg. Cost}}
        & \multicolumn{4}{c}{\textbf{$\mathrm{EX}_{\mathrm{col}}$ (\%)}} \\
        \cmidrule(lr){2-3}
        \cmidrule(lr){4-7}
        & \textbf{Tokens} & \textbf{CQs}
        & $\mathbf{T1}$ & $\mathbf{T2}$ & $\mathbf{T3}$ & $\mathbf{T4}$ \\
        \midrule
        \ours
        & 5.01K & 2.33
        & \textbf{70.66} & \textbf{74.05} & \textbf{76.21} & \textbf{77.38} \\
        -- w/o Graph
        & 3.55K & 1.56
        & 65.51 & 69.69 & 71.12 & 73.08 \\
        \makecell[l]{-- rp/w Query-Time}
        & 12.25K & 2.21
        & 69.23 & 72.56 & 74.58 & 75.81 \\
        \bottomrule
    \end{tabular}
\end{table}

Table~\ref{tab:graph-effectiveness} shows the results. 
The ambiguity graph substantially improves database-grounded ambiguity detection. Compared with w/o Graph, \ours gains 5.15\% at T1 and retains a 4.30\% advantage at T4, showing that many useful competing interpretations are difficult to rediscover through LLM reasoning alone. rp/w Query-Time closes part of this gap but remains 1.43\% below \ours at T1 and 1.57\% below at T4, while consuming 2.44$\times$ as many T1 LLM tokens. The lower accuracy indicates that, even when conditioned on the target NL question, reasoning over the full schema is less effective at identifying ambiguous column alternatives than offline pairwise reasoning over rich column profiles. rp/w Query-Time produces a similar number of CQs to \ours (2.21 vs.\ 2.33), indicating that the additional cost primarily comes from repeatedly searching the full schema rather than from increased user interaction.

We further examine the retrieved alternatives for each column anchor. On average, \emph{Graph-Based Retrieval} provides 4.80 candidate alternatives per anchor, whereas rp/w Query-Time returns only 2.49. Among them, 1.94 alternatives per anchor are shared by both settings; \emph{Graph-Based Retrieval} additionally contributes 2.86 alternatives missed by rp/w Query-Time, whereas rp/w Query-Time contributes only 0.55 alternatives absent from \emph{Graph-Based Retrieval}. This asymmetric coverage further suggests that offline pair-level reasoning over richer database-grounded profiles identifies a broader set of reusable competing alternatives than repeatedly searching the full schema for ambiguous columns at query time.

\subsubsection{Effectiveness of Multi-View Candidate Proposal}
\label{sec:construction-ablation}

We study the contribution and efficiency of the three candidate-proposal channels. We individually remove the \emph{schema-semantic}, \emph{surface-name}, and \emph{value-collision} channels from the complete \ours. We also compare with \textbf{All-Pairs Verification}, which bypasses candidate proposal and sends the exhaustive pairs to the same verifier. 

\begin{table}[h]
    \centering
    \caption{Effectiveness of Multi-view Candidate Proposal.}
    \label{tab:construction-ablation}
    \small
    \setlength{\tabcolsep}{4pt}
    \begin{tabular}{@{}l c c c c c c@{}}
        \toprule
        \textbf{Setting}
        & \makecell{\textbf{Verified}\\\textbf{Pairs}}
        & \makecell{\textbf{Accepted}\\\textbf{Pairs}}
        & \multicolumn{4}{c}{\textbf{$\mathrm{EX}_{\mathrm{col}}$ (\%)}} \\
        \cmidrule(lr){4-7}
        &
        {\scriptsize (Reduction)}
        & {\scriptsize (Avg. Degree)}
        & $\mathbf{T1}$
        & $\mathbf{T2}$
        & $\mathbf{T3}$
        & $\mathbf{T4}$ \\
        \midrule

        \makecell[l]{All\\Channels}
        & \makecell{6,276\\{\scriptsize(85.19\%)}}
        & \makecell{2,272\\{\scriptsize(5.69)}}
        & \textbf{70.66}
        & \textbf{74.05}
        & \textbf{76.21}
        & \textbf{77.38} \\

        \midrule

        \makecell[l]{w/o Schema-\\Semantic}
        & \makecell{3,525\\{\scriptsize(91.68\%)}}
        & \makecell{1,442\\{\scriptsize(3.61)}}
        & 69.49
        & 73.14
        & 75.23
        & 76.53 \\

        \makecell[l]{w/o Surface-\\Name}
        & \makecell{4,970\\{\scriptsize(88.27\%)}}
        & \makecell{1,929\\{\scriptsize(4.83)}}
        & 69.88
        & 73.73
        & 75.81
        & 76.47 \\

        \makecell[l]{w/o Value\\Collision}
        & \makecell{5,680\\{\scriptsize(86.60\%)}}
        & \makecell{2,195\\{\scriptsize(5.50)}}
        & 69.95
        & 73.86
        & 75.75
        & 76.66 \\

        \midrule

        All-Pairs
        & \makecell{42,387\\{\scriptsize(0.00\%)}}
        & \makecell{3,064\\{\scriptsize(7.68)}}
        & 70.01
        & 73.66
        & 75.88
        & 76.73 \\

        \bottomrule
    \end{tabular}
\end{table}

Table~\ref{tab:construction-ablation} shows the results.
The three channels exhibit overlap but remain complementary. Among the candidates proposed by each channel, 63.2\% of schema-semantic candidates and approximately half of the surface-name and value-collision candidates are unique to their respective views. Removing the schema-semantic, surface-name, and value-collision channels reduces the number of accepted pairs relative to using all channels by 36.5\%, 15.1\%, and 3.4\%, respectively. Removing any single channel also leads to a consistent downstream degradation of 0.72--0.91\% at T4, confirming the complementary contribution of the three channels.

Multi-view proposal also substantially reduces offline construction cost. Relative to All-Pairs Verification, it eliminates 85.19\% of pair-verification calls and produces a graph that is 25.8\% sparser in accepted relations (average degree 5.69 vs.\ 7.68). Nevertheless, it improves T4 accuracy from 76.73\% to 77.38\%. Thus, the proposal channels serve not only as an efficiency filter but also focus verification on more useful database-grounded relations, avoiding unnecessary graph connectivity without sacrificing downstream coverage.

\subsubsection{Effectiveness of Ambiguity Trigger Contexts}
\label{sec:trigger-context-ablation}

We study whether the ambiguity trigger contexts distilled offline provide useful guidance for ambiguity detection by comparing the full \ours with a variant that removes this information. We use turn-wise $\mathrm{EX}_{\mathrm{col}}$ to evaluate its downstream effectiveness.

\begin{table}[h]
    \centering
    \caption{Effectiveness of ambiguity trigger contexts.}
    \label{tab:trigger-context-ablation}
    \small
    \setlength{\tabcolsep}{4.5pt}
    \begin{tabular}{@{}l c c c c@{}}
        \toprule
        \textbf{Setting}
        & \multicolumn{4}{c}{\textbf{$\mathrm{EX}_{\mathrm{col}}$ (\%)}} \\
        \cmidrule(lr){2-5}
        & $\mathbf{T1}$ & $\mathbf{T2}$ & $\mathbf{T3}$ & $\mathbf{T4}$ \\
        \midrule
        \ours
        & \textbf{70.66} & \textbf{74.05} & \textbf{76.21} & \textbf{77.38} \\
        -- w/o Trigger Context
        & 68.84 & 72.69 & 74.38 & 75.75 \\
        \bottomrule
    \end{tabular}
\end{table}

As shown in Table~\ref{tab:trigger-context-ablation}, removing the trigger contexts decreases T1 $\mathrm{EX}_{\mathrm{col}}$ from 70.66\% to 68.84\%, and the gap remains 1.63\% at T4 (77.38\% vs.\ 75.75\%). Similar degradation is observed at the intermediate turns. These results indicate that graph connectivity alone is insufficient: explicitly describing \emph{when} two database columns may become competing interpretations helps the detector determine whether this relation is relevant to the current NL question.

\subsubsection{Effectiveness of Cross-Channel Integration}
\label{sec:integration-ablation}

We study whether Cross-Channel Integration (\ref{sec:cross-channel-integration}) improves the efficiency and effectiveness of ambiguity detection. We compare the full \ours with a variant that disables integration. We report T1 LLM cost, CQs per case, and choices per CQ to characterize cost, together with turn-wise $\mathrm{EX}_{\mathrm{col}}$ to measure downstream effectiveness.

\begin{table}[h]
    \centering
    \caption{Effectiveness of Cross-Channel Integration.}
    \label{tab:integration-ablation}
    \small
    \setlength{\tabcolsep}{2.2pt}
    \begin{tabular}{@{}l c c c c c c c@{}}
        \toprule
        \textbf{Setting}
        & \multicolumn{3}{c}{\textbf{T1 Cost}}
        & \multicolumn{4}{c}{\textbf{$\mathrm{EX}_{\mathrm{col}}$ (\%)}} \\
        \cmidrule(lr){2-4}
        \cmidrule(lr){5-8}
        & \makecell{\textbf{Tokens}\\\textbf{/Case}}
        & \makecell{\textbf{CQs}\\\textbf{/Case}}
        & \makecell{\textbf{Choices}\\\textbf{/CQ}}
        & $\mathbf{T1}$
        & $\mathbf{T2}$
        & $\mathbf{T3}$
        & $\mathbf{T4}$ \\
        \midrule

        \ours
        & 5.01K & 2.33 & 2.49
        & \textbf{70.66} & \textbf{74.05} & \textbf{76.21} & \textbf{77.38} \\

        -- w/o Integration
        & 5.14K & 2.36 & 2.50
        & 69.43 & 73.08 & 74.84 & 76.01 \\


        \bottomrule
    \end{tabular}
\end{table}

As shown in Table~\ref{tab:integration-ablation}, the full \ours reduces T1 token usage from 5.14K to 5.01K and CQs per case from 2.36 to 2.33, while the number of choices per CQ remains nearly unchanged. More importantly, integration improves accuracy at every turn, including gains of 1.23\% at T1 and 1.37\% at T4.

We further analyze the effect of integration using intermediate statistics. By comparing the column and value retrieval entries $(c,\mathcal{R}_{G}(c))$ and $((c,v),\mathcal{R}_{S}(c,v))$ before integration with those retained in $\mathcal{B}(q_0)$, we can quantify the pruning effect at two levels: the number of alternatives removed for each anchor (i.e., the number of items in $\mathcal{R}_{G}(c)$ or $\mathcal{R}_{S}(c,v)$), and the number of entire retrieval entries removed when no competing alternative remains. 
We observe that 1.33 alternatives per anchor on average are pruned, as well as 501 retrieval entries, corresponding to 0.69 entries per case.
If we conceptually consider that an entry with several retrieved alternatives may lead to CQs whose choices cover some or all of these alternatives, the reduction rate in CQs and choices is more modest than that of retrieval alternatives and entries. This can be attributed to the capability of LLM contextualization to exclude some irrelevant alternatives and unnecessary CQs through reasoning. 
Nevertheless, the consistent accuracy gains suggest that integration improves the quality of the information presented to the detector by removing unsupported alternatives. This provides more focused context for subsequent ambiguity reasoning.

\subsubsection{LLM Sensitivity Analysis}
\label{sec:llm-sensitivity}

Beyond the ablations of important components, we examine whether \ours is sensitive to the LLMs used in its offline construction and online disambiguation stages. 

\paragraph{Offline Construction Model}
\label{sec:llm-sensitivity-offline}

\begin{sloppypar}
Our main configuration uses \offlinemodel \cite{Qwen37Plus} for trigger-context induction and pair verification. Since it is a large proprietary model, we additionally evaluate Qwen3.8-27B \cite{AlibabaModelStudio}, a large open-source model from the same Qwen series. Both models verify the same candidate pairs produced by the full multi-view proposal pipeline, and the resulting graphs are evaluated using the same Qwen3.5-27B online pipeline.
\end{sloppypar}

\begin{table}[h]
    \centering
    \caption{Sensitivity to the offline construction model.}
    \label{tab:offline-llm-sensitivity}
    \small
    \setlength{\tabcolsep}{2.2pt}
    \begin{tabular}{@{}l c c c c c c@{}}
        \toprule
        \textbf{Model}
        & \makecell{\textbf{Accepted}\\\textbf{Pairs}}
        & \makecell{\textbf{Avg.}\\\textbf{Degree}}
        & \multicolumn{4}{c}{\textbf{$\mathrm{EX}_{\mathrm{col}}$ (\%)}} \\
        \cmidrule(lr){4-7}
        & & &
        $\mathbf{T1}$ & $\mathbf{T2}$ & $\mathbf{T3}$ & $\mathbf{T4}$ \\
        \midrule
        Qwen3.7-Plus
        & 2,272
        & 5.69
        & \textbf{70.66}
        & \textbf{74.05}
        & \textbf{76.21}
        & \textbf{77.38} \\
        Qwen3.8-27B
        & 916
        & 2.30
        & 69.30
        & 72.56
        & 74.64
        & 75.68 \\
        \bottomrule
    \end{tabular}
\end{table}

Qwen3.8-27B produces a substantially sparser graph, accepting 916 pairs compared with 2,272 for \offlinemodel, and consequently yields lower downstream accuracy. Nevertheless, its T4 accuracy reaches 75.68\%, only 1.70\% below the main configuration despite accepting 59.7\% fewer pairs. The performance gap is relatively modest under the large difference in graph density. One possible explanation is that Qwen3.8-27B is either less capable of inducing realistic trigger contexts or more conservative in accepting candidate pairs, resulting in lower graph coverage, whereas \offlinemodel accepts substantially more relations, some of which may provide limited downstream utility or introduce noise. These two effects may partially offset each other, making the downstream performance gap modest. Overall, these results suggest that offline construction quality is affected by verifier capability, while \ours remains effective when constructed with an open-weight model.

\paragraph{Online Disambiguation Model.}
\label{sec:llm-sensitivity-online}

We next vary the online reasoning model using the dense Qwen3.5 family~\cite{qwen35}, comparing the 4B, 9B, and 27B models for both the first turn and the subsequent \sphinteract continuation. The seed SQL is generated as introduced in Section~\ref{sec:implement-details} and is kept identical across all settings. \onlinemodel is fixed for User simulation, so changing the model size affects only the system-side LLM calls used for interactive disambiguation rather than the simulated user feedback. For each model size, we rerun \sphinteract~\cite{sphinteract} using the same online LLM and compare against its fourth turn, whose cumulative LLM cost is closest to the final cost of \ours across all settings.

\begin{table}[h]
    \centering
    \caption{Sensitivity to the online disambiguation model. Values in parentheses denote the average cumulative LLM cost per case.}
    \label{tab:online-llm-sensitivity}
    \small
    \setlength{\tabcolsep}{4pt}
    \begin{tabular}{@{}l c c c c c@{}}
        \toprule
        & \textbf{\sphinteract $\mathrm{EX}_{\mathrm{col}}$ (\%)}
        & \multicolumn{4}{c}{\textbf{\ours $\mathrm{EX}_{\mathrm{col}}$ (\%)}} \\
        \cmidrule(lr){2-2}
        \cmidrule(lr){3-6}
        \textbf{Size}
        & \makecell{\textbf{T4}}
        & $\mathbf{T1}$ & $\mathbf{T2}$ & $\mathbf{T3}$
        & \makecell{$\mathbf{T4}$} \\
        \midrule
        27B
        & 65.78 (10.20K)
        & 70.66 & 74.05 & 76.21 & \textbf{77.38} (9.78K) \\
        9B
        & 61.86 (10.60K)
        & 64.34 & 66.75 & 67.54 & \textbf{68.06} (10.59K) \\
        4B
        & 57.63 (12.35K)
        & 60.23 & 61.86 & 62.32 & \textbf{62.71} (12.19K) \\
        \bottomrule
    \end{tabular}
\end{table}

\ours consistently outperforms the corresponding \sphinteract baseline across all three model sizes while using comparable or slightly lower LLM cost. At T4, the gains are 11.60\%, 6.20\%, and 5.08\% for 27B, 9B, and 4B, respectively. The performance gap nevertheless narrows as the online model becomes smaller. This degradation appears to be partially attributable to weaker ambiguity detection: the average number of T1 CQs decreases from 2.33 with 27B to 1.85 with 9B and 1.71 with 4B, suggesting that lower-capacity models fail to surface some ambiguities even when provided with the same database-grounded guidance.
At the same turn, smaller models also incur higher cumulative LLM cost. This is because they successfully correct fewer cases in earlier turns, causing a larger fraction of cases to continue into subsequent turns and therefore increasing the average cumulative cost.

Notably, the database-grounded guidance partially compensates for reduced model capacity across scales. \ours with 9B reaches 68.06\%, exceeding the 65.78\% achieved by \sphinteract with 27B, while \ours with 4B reaches 62.71\%, also surpassing \sphinteract with 9B at 61.86\%. Although these cross-scale comparisons do not replace the controlled same-scale results, they indicate that explicit database-grounded ambiguity guidance can reduce the dependence of interactive disambiguation on a larger online LLM.

\subsubsection{Evaluation with BIRD Evidence}
\label{sec:with-bird-evidence}

Our main experiments exclude the evidence provided by BIRD~\cite{BIRD}. For completeness, we additionally evaluate the setting in which BIRD evidence is available. The evidence is provided to all system-side components, including seed SQL generation, ambiguity detection, and SQL regeneration, but not to the user simulator, which already has access to the gold SQL. Following Section~\ref{sec:llm-sensitivity-online}, we evaluate Qwen3.5-27B, 9B, and 4B~\cite{qwen35} as the online disambiguation model while fixing the user simulator to Qwen3.5-27B. The seed SQL is generated by the same Text-to-SQL parser introduced in Section~\ref{sec:implement-details} but with the benchmark-native evidence provided, and is kept identical across all settings. We compare \ours with \sphinteract~\cite{sphinteract} at T4, where the cost is closest to the final cost of \ours.

\begin{table}[h]
    \centering
    \caption{Evaluation with BIRD evidence. The same seed result is used across all settings, which achieves \textbf{65.71\%} $\mathrm{EX}_{\mathrm{col}}$. Values in parentheses denote average cumulative LLM cost per case.}
    \label{tab:with-bird-evidence}
    \small
    \setlength{\tabcolsep}{4pt}
    \begin{tabular}{@{}l c c c c c@{}}
        \toprule
        & \textbf{\sphinteract $\mathrm{EX}_{\mathrm{col}}$ (\%)}
        & \multicolumn{4}{c}{\textbf{\ours $\mathrm{EX}_{\mathrm{col}}$ (\%)}} \\
        \cmidrule(lr){2-2}
        \cmidrule(lr){3-6}
        \textbf{Size}
        & \textbf{T4}
        & $\mathbf{T1}$ & $\mathbf{T2}$ & $\mathbf{T3}$ & $\mathbf{T4}$ \\
        \midrule
        27B
        & 77.57 (8.42K)
        & 78.16 & 80.96 & 82.46 & \textbf{83.18} (8.48K) \\
        9B
        & 72.62 (9.10K)
        & 74.51 & 76.01 & 76.60 & \textbf{76.73} (9.44K) \\
        4B
        & 70.34 (9.80K)
        & 72.16 & 73.01 & 73.40 & \textbf{73.47} (9.92K) \\
        \bottomrule
    \end{tabular}
\end{table}

\ours consistently outperforms the same-scale \sphinteract baseline across all three model sizes under comparable system LLM cost. At T4, the gains are 5.61\%, 4.11\%, and 3.13\% for 27B, 9B, and 4B, respectively. These improvements are smaller than the corresponding gains in our main setting where evidence is not available. This reduction is expected because BIRD evidence supplies additional intent specifications and therefore resolves part of the ambiguity that would otherwise need to be identified through interaction. Nevertheless, the consistent advantage of \ours shows that database-grounded guidance remains useful even when the evidence is available. As in the evidence-free setting, smaller online models also ask fewer first-turn CQs (2.52, 2.04, and 1.75 for 27B, 9B, and 4B), suggesting that reduced ambiguity-detection capability remains one potential factor limiting lower-capacity models.

\section{Conclusion}
\label{sec:conclusion}

We presented \ours, a database-grounded framework for interactive ambiguity resolution in Text-to-SQL. \ours separates reusable ambiguity discovery from query-time reasoning by constructing a database-specific ambiguity graph offline and retrieving graph- and value-based guidance online. Experiments show that \ours significantly improves the accuracy--cost trade-off over existing interactive baselines, particularly in early clarification turns. These results demonstrate the value of database-grounded guidance for reliable and efficient interactive Text-to-SQL. Future work will extend \ours to dynamic databases, where schema or data changes may invalidate the ambiguity graph. Incremental maintenance could update affected relations without reconstructing the full graph.

\clearpage
\newpage

\bibliographystyle{ACM-Reference-Format}
\bibliography{ambigraph_refs}

\end{document}